\documentclass{article}
\usepackage{multirow}
\usepackage{spconf,amsmath,graphicx}
\usepackage{kotex}
\usepackage{booktabs}
\usepackage{comment}
\title{Experimental Study: Enhancing Voice Spoofing Detection Models with wav2vec 2.0}

\name{
    Taein Kang$^{1}$, Soyul Han$^{1}$, Sunmook Choi$^{2}$, Jaejin Seo$^{1}$, Sanghyeok Chung$^{2}$}
\names{
    Seungeun Lee$^{2}$, Seungsang Oh$^{2}$, Il-Youp Kwak$^{1}$
\thanks{This work was supported by the National Research Foundation of Korea (NRF) grant funded by the Ministry of Science and ICT (RS-2023-00208284) and Institute for Information \& communications Technology Planning \& Evaluation (IITP) grant funded by the Korea government(MSIT) (No.2019-0-00033, 50\%, Study on Quantum Security Evaluation of Cryptography based on Computational Quantum Complexity).}}
\address{$^{1}$ Chung-Ang University, Department of Statistics and Data Science, Republic of Korea \\ $^{2}$ Korea University, Department of Mathematics, Republic of Korea}

\begin{document}
%\ninept
%
\maketitle

\begin{abstract}
%Traditional spoofing detection systems have heavily relied on compressed handcrafted features of speech data. However, recently, there has been a notable paradigm shift towards directly utilizing the raw waveforms of speech data. This shift is exemplified by approaches like SincNet filters and emphasizes the growing demand for more sophisticated audio sample features. Additionally, the success of wav2vec 2.0-based models in spoofing detection has underscored the importance of refined feature encoders. In response, our research delved into assessing the representational capabilities of wav2vec 2.0 as audio features by exploring the number of frozen and trainable transformer layers. This analysis was conducted in comparison with traditional handcrafted features, and we have discovered models that achieve state-of-the-art performance in the ASVspoof2019 LA dataset. Through this comprehensive examination, our study offers insights into effective feature selection strategies for enhancing spoofing detection mechanisms.
% 오승상 교수님 
Conventional spoofing detection systems have heavily relied on the use of handcrafted features derived from speech data. However, a notable shift has recently emerged towards the direct utilization of raw speech waveforms, as demonstrated by methods like SincNet filters. This shift underscores the demand for more sophisticated audio sample features. Moreover, the success of deep learning models, particularly those utilizing large pretrained wav2vec 2.0 as a featurization front-end, highlights the importance of refined feature encoders. In response, this research assessed the representational capability of wav2vec 2.0 as an audio feature extractor, modifying the size of its pretrained Transformer layers through two key adjustments: (1) selecting a subset of layers starting from the leftmost one and (2) fine-tuning a portion of the selected layers from the rightmost one. We complemented this analysis with five spoofing detection back-end models, with a primary focus on AASIST, enabling us to pinpoint the optimal configuration for the selection and fine-tuning process. In contrast to conventional handcrafted features, our investigation identified several spoofing detection systems that achieve state-of-the-art performance in the ASVspoof 2019 LA dataset. This comprehensive exploration offers valuable insights into feature selection strategies, advancing the field of spoofing detection.

\end{abstract}

\section{Introduction}
\label{sec:intro}
In conventional spoofing systems, machine learning models, such as Gaussian mixture models (GMM), have typically been combined with handcrafted speech features. These handcrafted features, including constant Q cepstral coefficients (CQCC), Mel-frequency cepstral coefficients (MFCC), and linear frequency cepstral coefficients (LFCC), serve to condense spectral information and address challenges posed by highly correlated features. With the advent of deep learning, there has been a notable shift towards leveraging audio features in the spectro-temporal domain, resembling raw data more closely. These two-dimensional features, including constant-Q transform (CQT), short-time Fourier transform (STFT), and mel-spectrogram features, have become prevalent choices as inputs for spoofing countermeasure systems, as demonstrated by LCNN~\cite{lcnn, asv19la_cqt}.

Nevertheless, the selection of handcrafted features can be a burdensome task due to the numerous involved hyperparameters. In response, models like RawNet2~\cite{tak2021end} have emerged, taking raw waveforms as input and applying SincNet filters~\cite{sincnet} to achieve high spectral resolution. Similarly, RawGAT-ST~\cite{tak2021end2}, AASIST~\cite{jung2022aasist}, and the model for spectro-temporal dependency~\cite{gcn} utilize graph neural networks (GNN) and graph convolutional networks (GCN) to capture spectro-temporal relationships. 
Furthermore, the rise of large pretrained models in the speech domain, such as wav2vec 2.0 \cite{wav2vec2, babu2021xls}, WavLM \cite{chen2022wavlm}, and Hubert~\cite{hsu2021hubert} has garnered significant attention. In particular, the work by \cite{w2v2asp} focuses on countering synthetic voice attacks in speaker verification systems, exploring effective feature spaces and architectures using wav2vec 2.0. This approach has yielded significant performances in both spoof detection and spoofing-aware speaker verification tasks. 

%WavLM\cite{chen2022wavlm} 

\begin{figure}[th!]
\centering
\centerline{\includegraphics[width=8.5cm] % 8.5  12.5
{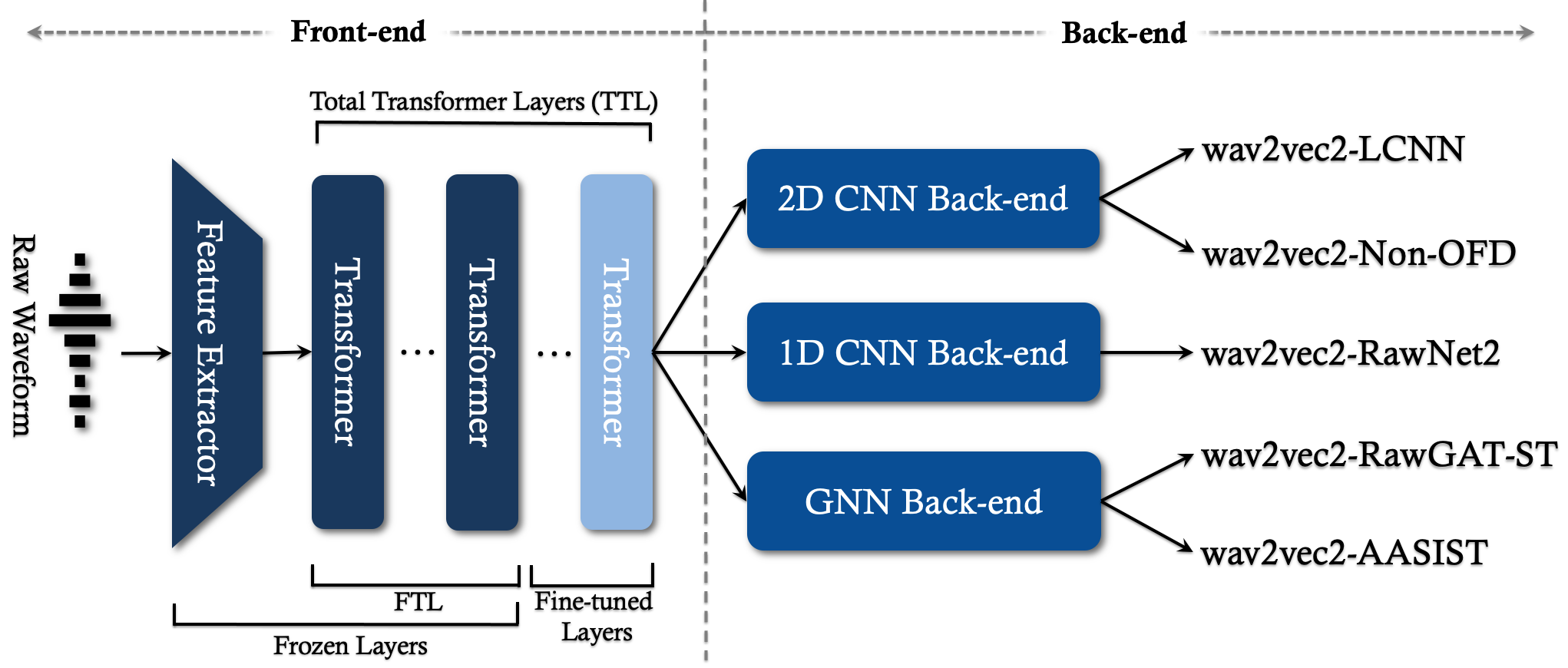}}
\caption{Proposed frameworks: pretrained wav2vec-based model as a featurization front-end, with the flexibility to fine-tune several Transformer layers, and followed by five distinct spoofing detection back-ends.
 %Proposed frameworks: pretrained wav2vec 2.0 model as a featurization front-end, with the flexibility to fine-tune a few Transformer layers, followed by five distinct spoofing detection models.
 }
\label{fig:system}
\end{figure}

This study primarily focuses on identifying the optimal wav2vec 2.0 model for utilization as an audio feature extractor in spoofing countermeasure systems. We conduct a comprehensive investigation into wav2vec 2.0, specifically determining the ideal number of pretrained transformer layers for feature extraction. Pretrained models often lack domain adaptation tailored to specific datasets. To tackle this challenge, we investigate fine-tuning strategies for the selected layers, aiming to enhance the model's adaptability to our specific data. With wav2vec 2.0 as our front-end, we employ five spoofing detection models, including LCNN, Non-OFD \cite{ofd}, RawNet2, RawGAT-ST and AASIST. Our in-depth analysis of the AASIST model significantly contributes to the identification of the appropriate transformer layer selection for wav2vec 2.0. This extensive examination provides valuable insights into feature selection strategies, ultimately enhancing spoofing detection. Consequently, this study presents an end-to-end training approach that demonstrates state-of-the-art performance when applied to the ASVspoof 2019 LA dataset.

\section{Methods}
\label{sec:methods}

%This study focuses on optimizing wav2vec-based models for audio feature extraction tailored for spoofing countermeasures. We carefully choose pretrained Transformer layers and fine-tune selected rightmost layers (shown in Figure \ref{fig:system}) to address the issue of domain adaptation tailored to our datasets. With pretrained and selectively fine-tuned wav2vec 2.0 as a front-end, we employ five distinct spoofing detection models as our back-ends, with a particular focus on the AASIST model.

As illustrated in Figure \ref{fig:system}, this study focuses on optimizing wav2vec-based models for spoofing countermeasures. We carefully choose the number of leftmost layers among the complete Transformer layers and fine-tune some rightmost layers within the selected ones to address the issue of domain adaptation tailored to our datasets. With pretrained and selectively fine-tuned wav2vec 2.0 as a front-end, we employ five distinct spoofing detection models as our back-ends, with a particular focus on the AASIST model.

\subsection{Pretrained wav2vec 2.0: selection and fine-tuning Transformer layers}

Wav2vec 2.0~\cite{wav2vec2}, a self-supervised learning (SSL) model building upon the original wav2vec model~\cite{schneider2019wav2vec}, extracts speech representations from raw audio data. Its convolution encoder initiates the process by generating feature maps that capture phonemes from raw waveforms. Subsequently, the Transformer architecture produces contextualized representations, and the last quantization module generates quantized representations using multiple codebooks. This model learns speech representations by using a contrastive loss, capturing dependencies across latent representations without labels.

Recent advancements have introduced many cross-lingual pretrained wav2vec 2.0 models, significantly expanding the number of languages, model size, and training data \cite{ babu2021xls,conneau21_interspeech}. For instance, XLSR-53 was trained on 50K hours of unannotated public speech data across 53 languages, while XLS-R(0.3B), XLS-R(1B), and XLS-R(2B) leveraged an impressive amount of 436K hours across 128 languages. It is important to note that while XLSR-53 and XLS-R(0.3B) have 24 Transformer layers with the same 317M parameters, XLS-R(1B) and XLS-R(2B) boast 48 Transformer layers with 965M and 2162M parameters, respectively. These pretrained models are readily accessible via Hugging Face \cite{huggingface}. % and constituted a fundamental component of our experimental setup.

%Pretrained models often exhibit limitations in domain adaptation, which is a crucial necessity for effectively aligning with the unique characteristics of specific datasets.
Pretrained models often exhibit limitations in domain adaptation, making it crucial to align them with the unique characteristics of specific datasets.
In response to this pivotal challenge, our research meticulously explores fine-tuning strategies for selectively chosen Transformer layers. Previous studies have explored the utilization of wav2vec 2.0 as a feature extractor for spoofing detection \cite{w2v2asp, w2v2vib,  9747768}. 
Building upon these prior approaches, our research focuses on assessing the potential enhancements attainable within existing spoofing countermeasure back-ends through the adoption of wav2vec 2.0 features as audio representations.

Our research is underpinned by the hypothesis that a more tailored and effective feature extraction can be achieved in the realm of spoofing detection by fine-tuning specific pretrained parameters of Transformer layers within wav2vec-based models. In this context, we introduce two pivotal hyperparameters: the total number of Transformer layers (\#TTL) and the number of frozen Transformer layers (\#FTL). In our detailed implementation, we select the \#TTL in the leftmost Transformer layers from the entire architecture, designate the \#FTL in the left Transformer layers among the selected ones as frozen during the training process, and focus on fine-tuning the remaining \#TTL $\!-\!$ \#FTL in the rightmost Transformer layers. For our case study, we initiated this exploration with XLS-R(1B), aiming to discern the optimal hyperparameters \#TTL and \#FTL. A schematic overview of our experimental framework is illustrated in Figure \ref{fig:system}.

\subsection{Spoofing detection back-ends}\label{subsec:spoofing-detection-models}

In this section, we introduce five spoofing countermeasure systems that form the basis of our experimental investigations. LCNN and Non-OFD are grounded in 2D convolutional neural networks, specializing in processing STFT and CQT spectrograms within the spectro-temporal domain derived from raw audio data. This transformation facilitates the extraction of frequency-related features, enhancing the efficiency of spoofing detection models in extracting pertinent information from 2D inputs. In contrast, the remaining three models are designed to operate directly with raw waveforms. Acknowledging the inherent challenge of extracting high spectral information from raw audio inputs, these models employ 1D CNN-based SincNet filters as their front-ends. Subsequently, the output of the filters is fed into their own architecture. For example, the RawNet2 model consists of six residual blocks, each incorporating 1D convolutions. The RawGAT-ST and AASIST models leverage graph attention networks (GAT) \cite{gat}. 

Among these five models, we have selected AASIST as the baseline of our back-end model. Its residual encoder generates two heterogeneous graph modules, one for spectral and the other for temporal information, utilizing SincNet filters as a front-end. These modules are then combined, akin to the RawNet2-based encoder, and followed by a series of heterogeneous stacking graph attention layers and max graph operations. %In our study, we employ the AASIST back-end only, substituting SincNet filters with a wav2vec-based front-end.

\section{Experiment}
\label{sec:experiment}

\subsection{ASVspoof 2019 LA dataset}
%The ASVspoof 2019~\cite{ASVspoof2019} Logical Access (LA) dataset serves as an evaluation benchmark for Text-to-Speech (TTS) and Voice Conversion (VC) attacks. It includes logical access scenarios, features 19 advanced TTS and VC systems, and offers training, development, and evaluation subsets.
%It includes logical access scenarios, features 19 advanced TTS and VC systems. The training and development sets consist of 6 TTS and VC attacks, while the evaluation set comprises 2 known attacks, 4 partially known attacks, and 7 unknown attacks\cite{tak2023end}. The dataset encompasses with around 25,000 training and development utterances and approximately 71,000 test utterances.

The ASVspoof 2019 Logical Access (LA) dataset~\cite{ASVspoof2019} serves as an evaluation benchmark primarily designed for testing the robustness of Automatic Speaker Verification (ASV) systems against emerging spoofing attacks. In addition to its role in ASV, this dataset also functions as a valuable assessment platform for Text-to-Speech (TTS) and Voice Conversion (VC) attacks. It encompasses a wide range of logical access scenarios created by 19 advanced TTS and VC systems, all capable of generating synthetic or converted speech for spoofing purposes. This dataset offers distinct training, development, and evaluation subsets, facilitating rigorous testing and validation of spoofing countermeasure systems.

\subsection{Optimizing \#TTL and \#FTL with AASIST}
%\subsection{Experimental results}

%(실험부분 제가 어떤 실험들이 있는지 구체적으로 잘 모르니 태인학생이 일단 초안잡아서 써봐주세요)

\begin{table}[htb]
\resizebox{\linewidth}{!} {%
%\begin{tiny}
    \begin{tabular}{c|ccccccccc}
%    \hline \hline
    \toprule[1pt]
    {\#TTL} & \multicolumn{9}{c}{\#FTL} \\ \cline{2-10} 
    %& 0$^{\text{th}}$ & 3$^{\text{rd}}$ & 6$^{\text{th}}$ & 9$^{\text{th}}$ & 12$^{\text{th}}$ & 15$^{\text{th}}$ & 18$^{\text{th}}$  & 33$^{\text{rd}}$  & 48$^{\text{th}}$ \\ \midrule[1pt]%\hline
    & 0 & 3 & 6 & 9 & 12 & 15 & 18 & 33 & 48 \\ \midrule[1pt]%\hline
 3   & 8.96&38.23&  -  &  -  &  -  &  -  &  -   &  -  &  -  \\ \hline
 6  & 7.71& 6.54&42.82&  -  &  -  &  -  &  -   &  -  &  -  \\ \hline
 9   & 1.38& 0.88& 0.98&17.84&  -  &  -  &  -   &  -  &  -  \\ \hline
 12  & 0.77& 0.41& 0.57& \textbf{0.22}&17.85&  -  &  -  &  -  &  -  \\ \hline
 15  & 0.87& 1.04& 1.35& 0.80& 1.43&41.00&  -  &  -  &  -  \\ \hline
 18  & 0.63& 0.40& 0.61& 2.14& 2.39& 6.68&26.01&  -  &  -  \\ \hline
 33  & 2.41& 2.01& 3.94& 3.57& 4.26& 4.36& 4.54&37.51&  -  \\ \hline
 48  & 2.37& 5.75& 0.65& 2.83& 7.52& 5.13& 9.72& 4.53&41.09\\ \bottomrule[1pt]%\hline \hline
\end{tabular}% 
}
\caption{EER(\%) results on XLS-R(1B) and AASIST combination system. %FTL 0$^{\text{th}}$ transformer layers means that all transformer layers are trainable. %And if TTL equal Freezing layers, wav2vec2 is not fine-tuned. 
}
%\caption{Experimental results of the combination of XLS-R (1B) and AASIST on ASVspoof LA dataset}

\label{tab:param_ex}
%\end{tiny}
\end{table}

In our experiment, we integrated the XLS-R(1B) front-end with the AASIST back-end. Our optimization approach involved systematically varying the hyperparameters \#TTL and \#FTL across the entire span from 0 to 48, employing a 3-layer interval within the complete set of 48 Transformer layers in XLS-R(1B). For performance assessment, we employed two key metrics: the equal error rate (EER) and the minimum normalized tandem detection cost function (t-DCF). Our training comprised 15 epochs, and we evaluated the model based on the instance that produced the lowest EER on the development dataset. Our optimization leveraged the Adam optimizer with an initial learning rate of 5e-5. The learning process is implemented through the PyTorch framework.

Table \ref{tab:param_ex} presents the resulting EERs. When \#TTL equaled \#FTL (i.e., when all parameters remained frozen without fine-tuning), the model’s performance deteriorated compared to the use of SincNet filters. This observation provided a compelling motivation to fine-tune selected Transformer layers. Within the table, when \#TTL ranged from 9 to 18, with \#FTL being strictly less than \#TTL, a consistent trend showing superior performance emerged. Specifically, when \#TTL was set at 12, all fine-tuned configurations outperformed the original AASIST model employing SincNet filters, achieving an EER of 0.83\%. Notably, the model configured with (\#TTL, \#FTL) = (12, 9) achieved an outstanding EER of 0.22\%. This underscores the potential of wav2vec 2.0 in enhancing pre-existing spoofing detection models with its robust audio representation capabilities.

\subsection{Various wav2vec 2.0 front-ends and spoofing detection back-ends}
\label{subsec:wav2vec-versions}
%먼저 dataset에 맞는 feature를 추출하기 위해, 그리고 Wav2vec2.0의 어느 transformer layer가 spoofing detection에 잘 적합 하는 지 확인하기 위해, Wav2vec2.0의 transformer layer에 변화를 주어 ASVSpoof2019 LA dataset에 대해 실험했다. 

\begin{table}[ht]
\centering
\begin{tabular}{lcccc}
\toprule[1pt]%\hline
version     & \#TTL & \#FTL & min t-DCF & EER(\%) \\ \midrule[1pt]%\hline
XLSR-53    & 12 &  3  & 0.0083&  0.26   \\ \hline
XLS-R(0.3B) & 15 &  6  & 0.0093&  0.29   \\ \hline
XLS-R(1B)   & 12 &  9  & \textbf{0.0063}&  \textbf{0.22}   \\ \hline
XLS-R(2B)   & 18 &  3  & 0.0098&  0.30\\ \bottomrule[1pt]%\hline
\end{tabular}
\caption{Results on various wav2vec 2.0 front-ends. %TTL denotes that \# of Total Transformer Layers. FNL denotes that Freezing parameters upto $N^{th}$ Layers.
}
\label{tab:version}
\end{table}

We conducted an expanded experiment study involving a broader array of wav2vec 2.0 front-ends: XLSR-53, XLS-R(0.3B), and XLS-R(2B). To streamline our exploration, we focused on a subset of hyperparameter pairs (\#TTL, \#FTL) that had demonstrated notably robust performance in our earlier experiment with XLS-R(1B). Specifically, we restricted \#TTL to the set \{12, 15, 18\} and \#FTL to the set \{0, 3, 6, 9, 12, 15\}, ensuring that \#FTL was strictly less than \#TTL. Table \ref{tab:version} provides a concise summary of the outcomes, presenting the \#TTL and \#FTL combinations that yielded the most favorable minimum t-DCF and EER values for each front-end. Significantly, all models exhibited exceptional performance, affirming the effectiveness of our selection and fine-tuning approach across various wav2vec-based models.

\begin{table}[ht]
\resizebox{\linewidth}{!}{%
%\begin{small}
\begin{tabular}{llcc}

\toprule[1pt]
Model                                       & Front-end         & min t-DCF & EER(\%) \\ \midrule[1pt]
\multirow{2}{*}{LCNN~\cite{asv19la_cqt}}     & STFT              &  0.1028   &  4.53   \\ \cline{2-4}
                                            & XLS-R(1B) (12/9) &\textbf{0.0320}&\textbf{1.02}\\ \hline
\multirow{2}{*}{Non-OFD~\cite{ofd}}          & CQT               &  -        &  1.35   \\ \cline{2-4}
                                            & XLS-R(1B) (15/9) &0.0111&\textbf{0.41}\\ \hline
\multirow{2}{*}{RawNet2~\cite{tak2021end}}   & SincNet filter    &  0.1301   &  5.64   \\ \cline{2-4}
                                            & XLS-R(1B) (12/6) &\textbf{0.0032}&\textbf{0.12}\\ \hline
\multirow{2}{*}{RawGAT-ST~\cite{tak2021end2}}& SincNet filter    &  0.0335   &  1.06   \\ \cline{2-4}
                                            & XLS-R(1B) (18/12) &\textbf{0.0048}&\textbf{0.24}\\ \hline
\multirow{2}{*}{AASIST~\cite{jung2022aasist}}& SincNet filter    &  0.0275   &  0.83   \\ \cline{2-4}
                                            & XLS-R(1B) (12/9) &\textbf{0.0063}&\textbf{0.22}\\ \bottomrule[1pt]
\end{tabular}%
}
\caption{Results on various spoofing detection back-ends.}
\label{tab:feature}
%\end{small}
\end{table}

In a separate experiment, we explored different spoofing detection back-ends: LCNN, Non-OFD, RawNet2, RawGAT-ST and AASIST. Each back-end has experimented with two distinct front-ends: 1) the original model front-end, which includes options like SincNet filter, STFT, or CQT, and 2) the XLS-R(1B) front-end, which allows for selection and fine-tuning. The results are presented in Table \ref{tab:feature}, in which `XLS-R(1B) $(m/n)$’ indicates the optimal configuration for \#TTL $ =m$ and for \#FTL $ =n$. Across all systems, the incorporation of wav2vec 2.0 as a front-end consistently resulted in significant performance enhancements compared to the use of the original model front-end. Furthermore, we reconfirm the suitability of exploring \#TTL values within the range of \{12, 15, 18\} for spoofing detection in the ASVspoof 2019 LA dataset. Most notably, the XLS-R(1B) and RawNet2 system demonstrated remarkable performance improvements, achieving a state-of-the-art minimum t-DCF of 0.0032 and an EER of 0.12\%, to the best of our knowledge.

\section{Discussion}
\label{sec:discussion}
%\subsection{Performance of w2v2-aasist in ADD 2023 competition}
%대회 참여 내용과 w2v2-aasist, cqt-aasist성능 관해 소개.

%\subsection{추가 토론거리}
%있다면 작성

%%% 이렇게 씁시다. 
%우리는 2023년 5월에 International Joint Conferences on Artificial Intelligence (IJCAI) 학회의 competition으로 개최된 Audio Deepfake Detection Challenge 2023 (ADD 2023) Track1.2에 CAU\_KU 팀으로 참여하여 3등을 차지하였다. wav2vec2 pre-trained 모형은 우리 모형들에 중요한 부분을 차지하였고, 본 페이퍼에서는 대회 페이퍼에서 자세히 다루지 못한 wav2vec2의 하이퍼파라메터 튜닝과 실제 적용 및 여러 실험들에 대해 더 자세히 다루었다.
%\subsection{Challenge participation with wav2vec 2.0 pretrained network}
%In May 2023, we participated as the CAU\_KU team in Track1.2 of the Audio Deepfake Detection Challenge 2023 (ADD 2023), held as a competition during the International Joint Conferences on Artificial Intelligence (IJCAI) \cite{add2023}. We ranked the third position \cite{cauku}, and the wav2vec 2.0 pre-trained model played a crucial role in our models. Throughout this paper, we delved further into the hyperparameter-tuning, practical applications, and various experiments related to wav2vec 2.0 that were not covered in the competition paper.

In Track1.2 (audio fake game - detection task) of the Audio Deepfake Detection Challenge 2023 (ADD 2023), hosted during the International Joint Conferences on Artificial Intelligence (IJCAI) \cite{add2023}, our CAU\_KU team achieved a third-place ranking \cite{cauku}. %Our approach in the competition resembled the experimental setup outlined in this paper. 
We utilized a diverse array of front-ends, including MFCC, CQT and wav2vec 2.0, alongside various back-ends such as LCNN, GMM, AASIST and OFD. During the competition, we ensembled three systems, CQT + LCNN, CQT + AASIST, and wav2vec 2.0 + GMM. Among those single systems, the wav2vec 2.0 + GMM system achieved the best performance. 
%During the competition, one of our ensemble systems was the wav2vec 2.0 + GMM system. We employed the base version of pretrained wav2vec 2.0, which was pretrained on 960 hours of speech data~\cite{wav2vec2}, as a front-end. 
Motivated by our participation in the challenge, we contemplated how to better utilize wav2vec2, leading to the research presented in this paper.

%\subsection{Comparison with recent state-of-the-art systems}

\begin{table}[th]
\resizebox{\linewidth}{!}{%
%\begin{footnotesize}
\begin{tabular}{llcc}

\toprule[1pt]
Model                                       & Front-end         & min t-DCF & EER(\%) \\ \midrule[1pt]
RawNet2~\cite{tak2021end}   & SincNet filter    &  0.1301   &  5.64   \\ \hline
GAT-T~\cite{gatt}           & LFB               &  0.0894   &  4.71   \\ \hline
LCNN~\cite{asv19la_cqt}     & STFT              &  0.1028   &  4.53   \\ \hline
GMM~\cite{gmm}              & LFCC              &  0.0904   &  3.50   \\ \hline
RW-ResNet~\cite{rwresnet}   & Raw Waveform      &  0.0817   &  2.98   \\ \hline
LCNN-LSTM-sum~\cite{lls}    & LFCC              &  0.0524   &  1.92   \\ \hline
Non-OFD~\cite{ofd}          & CQT               &  -        &  1.35   \\ \hline
RawGAT-ST~\cite{tak2021end2}& SincNet filter    &  0.0335   &  1.06   \\ \hline
AASIST~\cite{jung2022aasist}& SincNet filter    &  0.0275   &  0.83   \\ \hline
GCN based model~\cite{gcn}  & LFB               &  0.0166   &  0.58   \\ \hline
wav2vec 2.0 + VIB~\cite{w2v2vib} & BASE~\cite{wav2vec2}  &  0.0107   &  0.40   \\  \hline
wav2vec 2.0 + ASP~\cite{w2v2asp} & XLSR-53 &  -        &  0.31   \\  \hline \hline
\textbf{wav2vec 2.0 + AASIST (Ours)}&\textbf{XLS-R(1B)}&\textbf{0.0063}&\textbf{0.22}\\ \hline
\textbf{wav2vec 2.0 + RawNet2 (Ours)}&\textbf{XLS-R(1B)}&\textbf{0.0032}&\textbf{0.12}   \\ 
  \bottomrule[1pt]
\end{tabular}%
}
\caption{Comparison with recently established spoofing detection systems.}
\label{tab:comparision}
%\end{footnotesize}
\end{table}
In our current research, we leveraged transfer learning principles to enhance domain adaptation tailored to audio speech datasets. This involved strategically reducing the number of Transformer layers and fine-tuning a select portion of these layers. Our approach systematically adjusted two critical hyperparameters: the number of layers selected (\#TTL) and the number of layers fine-tuned (\#FTL). The performance results of our proposed models, along with several other established spoofing detection systems, on the ASVspoof 2019 LA evaluation data are presented in Table \ref{tab:comparision}. Notably, the two wav2vec-based systems, wav2vec 2.0 + VIB (variational information bottleneck module back-end) and wav2vec 2.0 + ASP (attentive statistics pooling layer back-end), demonstrated excellent performance, affirming the efficacy of wav2vec 2.0 as a front-end (The wav2vec 2.0 + VIB model used the base version of pretrained wav2vec 2.0, pretrained on 960 hours of speech data~\cite{wav2vec2}, as a front-end, and the wav2vec 2.0 + ASP model utilized XLSR-53~\cite{babu2021xls} as its front-end.).
 Moreover, our fine-tuned models, wav2vec 2.0 + AASIST and wav2vec 2.0 + RawNet2, which were identified through a rigorous two-hyperparameter search, consistently outperformed other models. In the future, we intend to further investigate the applicability of our proposed method with other state-of-the-art spoofing detection models.

\section{Conclusion}
\label{sec:conclusion}

%본 연구에서는 wav2vec2.0을 front-end로 활용하여 spoofing detection에 적합한 transformer layer를 찾고, trainable transformer layer를 조절하며 데이터셋에 wav2vec2.0을 최적화시켰다. Wav2vec2.0 network에 훈련된 데이터가 다를 때, 훈련된 parameter 수가 다를 때의 버전들과 비교하며 모델의 버전에 따라 spoofing detection에 적합한 transformer layer가 같은 지 검증했다. 또한 Wav2vec2.0을 front-end 삼아, 여러 모델을 결합시켜 결과를 보았고 그 중 wav2vec2-RawNet2와 wav2vec2-AASIST는 각각 EER 0.12\%, 0.22\%를 기록하며 state-of-the-art의 성능을 달성했다. 새로운 데이터 셋에 대한 적합을 시킬 때 계산 비용과 시간이 많이 든다는 단점이 있지만 본 연구가 다른 연구자들의 연구의 비용을 줄이는데 도움이 되었으면 한다.

%In this study, we utilized wav2vec 2.0 as a front-end to identify transformer layers suitable for spoofing detection and fine-tuned the trainable transformer layers to optimize the wav2vec 2.0 model. We compared various versions of the model based on variations in trained data and the number of parameters when trained on different datasets, verifying if the same transformer layers are suitable for spoofing detection across model versions. Additionally, using wav2vec 2.0 as a front-end, we combined several models and observed the results. Among them, wav2vec 2.0 + RawNet2 and wav2vec 2.0 + AASIST achieved state-of-the-art performance, each recording EERs of 0.12\% and 0.22\%, respectively. While it is worth noting that adapting to new datasets can be computationally expensive and time-consuming, we hope that this study contributes to reducing computational costs for other researchers.

In summary, our research underscores a paradigm shift in spoofing countermeasure systems – transitioning from traditional handcrafted speech features towards the adoption of deep learning-based audio representations. Leveraging a large pretrained wav2vec 2.0 model as a potent feature extractor directly from raw waveforms, we meticulously selected and fine-tuned a subset of its complete Transformer layers to enhance its adaptability to specific datasets by introducing two new hyperparameters. Our exploration encompassed various versions of wav2vec 2.0, revealing that different pretrained models can efficiently serve as effective front-ends. We complemented this with diverse spoofing detection back-end models like RawNet2 and AASIST across different categories, allowing us pinpoint the optimal configuration for the selection and fine-tuning process applied to wav2vec 2.0 models. Notably, our efforts unveiled wav2vec 2.0 + RawNet2 and wav2vec 2.0 + AASIST as standout performers, achieving state-of-the-art results with EERs of 0.12\% and 0.22\%, respectively, on the ASVspoof 2019 LA dataset. While we acknowledge the computational resources and time investments required for adapting to new datasets, we believe this research will serve as a valuable resource, alleviating these challenges for fellow researchers in the field.

% In this study, we utilized wav2vec 2.0 as a front-end to identify whether pretrained models are suitable for spoofing detection. We fine-tuned some transformer layers by introducing two hyperparameters to optimize the wav2vec 2.0 model's performance. Our exploration encompasses various versions of wav2vec 2.0, showing that different pretrained models are also capable of providing excellent features as a front-end. Additionally, we showed that pretrained model is well-suited for other spoofing detection models in different categories. Remarkably, our efforts led to the discovery of wav2vec 2.0 + RawNet2 and wav2vec 2.0 + AASIST, both of which achieved state-of-the-art performance, achieving EERs of 0.12\% and 0.22\%, respectively. While acknowledging the computational costs and time associated with adapting to new datasets, we aspire that this study will alleviate such burdens for fellow researchers. %Our findings underscore the potential of wav2vec 2.0 as a powerful tool in the domain of spoofing detection, offering both efficiency and superior performance.

%\section{Acknowledgments}
%\label{sec:Acknowledgments}

%This work was supported by the National Research Foundation of Korea (NRF) grant funded by the Ministry of Science and ICT (RS-2023-00208284) and Institute for Information \& communications Technology Planning \& Evaluation (IITP) grant funded by the Korea government(MSIT) (No.2019-0-00033, 50\%, Study on Quantum Security Evaluation of Cryptography based on Computational Quantum Complexity).

\vfill\pagebreak

\bibliographystyle{IEEEbib}
\bibliography{refs}

\end{document}